\documentclass[lnicst,sechang]{svmultln}

\usepackage{amssymb}
\usepackage{graphicx}
\usepackage{bm}
\usepackage{bbm}
\usepackage{units}
\usepackage{algorithm2e}
\RestyleAlgo{ruled}
\usepackage{amsmath}
\usepackage[english]{babel}
\usepackage[autostyle]{csquotes}

\newcommand{\comment}[1]{}

\begin{document}
\mainmatter

\title{Network Coding Over SATCOM: Lessons Learned}
\titlerunning{Network Coding Over SATCOM: Lessons Learned}

\author{Jason Cloud, Muriel M\'edard}
\authorrunning{Cloud et al.}

\tocauthor{Jason Cloud, Muriel M\'edard}

\institute{Massachusetts Institute of Technology, Cambridge, MA 02139, USA,\\
jcloud@mit.edu, medard@mit.edu}

\maketitle

\begin{abstract}

Satellite networks provide unique challenges that can restrict users'
quality of service. For example, high packet erasure rates and large latencies
can cause significant disruptions to applications such as video streaming or
voice-over-IP. Network coding is one promising technique that has been shown to
help improve performance, especially in these environments. However,
implementing any form of network code can be challenging. This paper will use
an example of a generation-based network code and a sliding-window network code
to help highlight the benefits and drawbacks of using one over the other.
In-order packet delivery delay, as well as network efficiency, will be used as
metrics to help differentiate between the two approaches. Furthermore, lessoned
learned during the course of our research will be provided in an attempt to help
the reader understand when and where network coding provides its benefits.

\end{abstract} 

\section{Introduction} 
\label{sec:Introduction}

Space-based packet data networks are becoming a necessity in everyday life,
especially when considering world-wide Internet connectivity. It is estimated
that over half of the world's population still does not have access to broadband
Internet due to a variety of factors including a lack of infrastructure and low
affordability, especially in rural areas and developing countries
\cite{Internet_barriers}. To overcome these barriers, a number of companies such as
SpaceX, Google, and FaceBook have recently launched projects that incorporate
some form of space-based or high altitude data packet network. However,
significant challenges such as large latencies, high packet erasure rates, and
legacy protocols (e.g., TCP) can seriously degrade performance and inhibit the
user's quality of service. One promising approach to help in these challenged
environments is network coding. This paper will investigate some of the gains
that network coding provides, as well as outline some of the lessons learned
from our research.

Space-based networks have a number of unique
characteristics that challenge high
quality of service applications. Large packet latencies and relatively high
packet erasure rates can negatively impact existing protocols. Fading due to
scintillation or other atmospheric effects are more pronounced than in terrestrial networks. The high
cost in terms of both deployment and bandwidth make efficient communication a
requirement. Finally, the broadcast nature of satellite networks create unique
challenges that are non-existent in terrestrial networks. While existing
physical and
data link layer techniques help improve performance in these conditions, we will
show that coding above these layers can also provide performance gains.

\begin{figure}
	\begin{centering}
	\includegraphics[width=1.0\columnwidth]{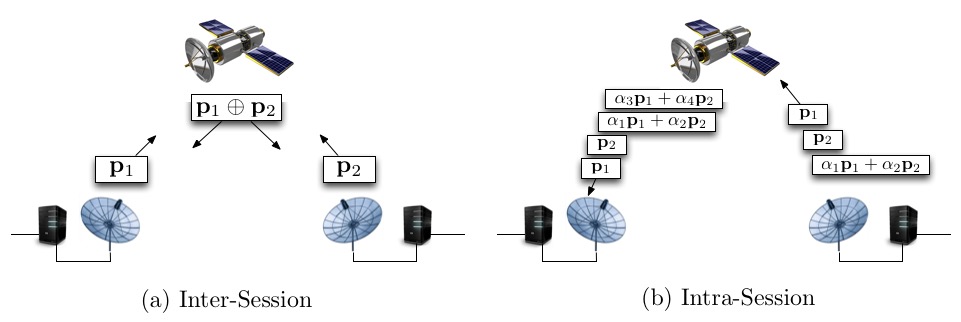}
	\par\end{centering}
	\caption{Examples of inter-session (a) and intra-session (b) network coding.
	This paper focuses completely on intra-session network coding.}
	\label{fig:NCType_Examples}
\end{figure}

Various forms of network coding can be used with great benefits in space-based
networks. In general, these can be characterized into two broad categories:
inter-session network coding, and intra-session network coding. Figure
\ref{fig:NCType_Examples} provides a simple example of both. Inter-session
network coding combines information flows together to improve the network
capacity. A summary of the various methods that can be used for satellite
communications is provided by Vieira \textit{et al.} \cite{5586880}. 
Intra-session network coding, on the other hand, is used to add redundancy into a
single information flow. Adding this redundancy has shown that file
transfer times can be decreased for both multicast \cite{6134279} and unicast
\cite{5062100,5513768} sessions.

While there are merits to both techniques, our focus will be on intra-session
network coding techniques that help achieve the following goals: provide
consistent performance for protocols not designed for space systems; decrease
delay for real-time or near real-time data streams; efficiently use any
network resources that are available; and reduce packet erasure rates due to
correlated losses. A generation-based approach
\cite{5513768,Cloud_CodedGenSRARQ_2015} and a sliding-window approach
\cite{Karzand_LDRLC_2014} will be used to help highlight the potential gains,
design choices, and implementation decisions that need to be taken into
account. Several performance metrics including the in-order delivery delay,
efficiency, and upper layer packet erasure rates will be used to help
differentiate between the approaches.

The remainder of the paper is organized as follows. Section 
\ref{sec:Implementing_NC} will provide details on the coding algorithms
considered. Section \ref{sec:performance_of_nc} provides information about the
assumed network model and evaluates the performance of these coding algorithms
when used for both reliable and unreliable data streams. Section 
\ref{sec:Implementation_Considerations} discusses various considerations that
need to be taken into account when implementing network coding into real
systems. Finally, conclusions are summarized in Section \ref{sec:Conclusion}.

\section{Network Coding over Packet Streams}
\label{sec:Implementing_NC}

Network coding has been shown to dramatically improve network
performance; however, implementing it can be a challenge. In order to develop
practical coding techniques, random linear network coding (RLNC) \cite{Ho_RLNC_2006}
has been used by a large number of coding schemes because of its simplicity and
effectiveness in most network scenarios. While both practical inter and
intra-session techniques have been proposed, we are primarily
interested in the latter due to the inherent limitations of existing satellite
communication networks (i.e., typical satellite communication networks employ a
bent-pipe architecture or have very limited on-orbit processing power). Assume
that we want to send a file consisting of information packets $\mathbf{p}_i$,
$i\in \mathcal{P}$, where $\mathcal{P}$ is the set of information packet
indexes (i.e., the file has size $\left|\mathcal{P}\right|$ packets). Within
these
intra-session packet streams, RLNC can be used to add redundancy
by treating each $\mathbf{p}_i$ as a vector in some finite field
$\mathbb{F}_{2^q}$. Random coefficients $\alpha_{ij}\in \mathbb{F}_{2^q}$ are
chosen, and linear combinations of the form $\mathbf{c}_i=\sum_{j\in
\mathcal{P}} \alpha_{ij} \mathbf{p_j}$ are generated. These coded packets are
then inserted at strategic locations to help overcome packet losses in lossy
networks.

Management of the coding windows for these intra-session network coding schemes
generally fall within the following two categories:
fixed-length/generation-based schemes, or variable/sliding window based
schemes. Fixed-length or generation-based schemes first partition information
packets into blocks, or generations, $G_i=\left\{\mathbf{p}_{
\left(i-1\right)k+1},\ldots,\mathbf{p}_{\min\left(ik,
\left|\mathcal{P}\right|\right)}\right\}$ for $i=\left[1,\left\lceil \nicefrac{
\left|\mathcal{P}\right|}{k}\right\rceil\right]$ and generation size $k\geq 1$.
Coded packets are then produced based on the information packets contained
within each individual generation. As a result, coded packets consisting of
linear combinations of packets in generation $G_i$ cannot be used to help decode
generation $G_j,i\neq j$. Alternatively, sliding window schemes do not impose
this restriction. Instead, information packets are dynamically included or
excluded from linear combinations based on various performance requirements.

\begin{figure}
\begin{centering}
\includegraphics[width=1.0\columnwidth]{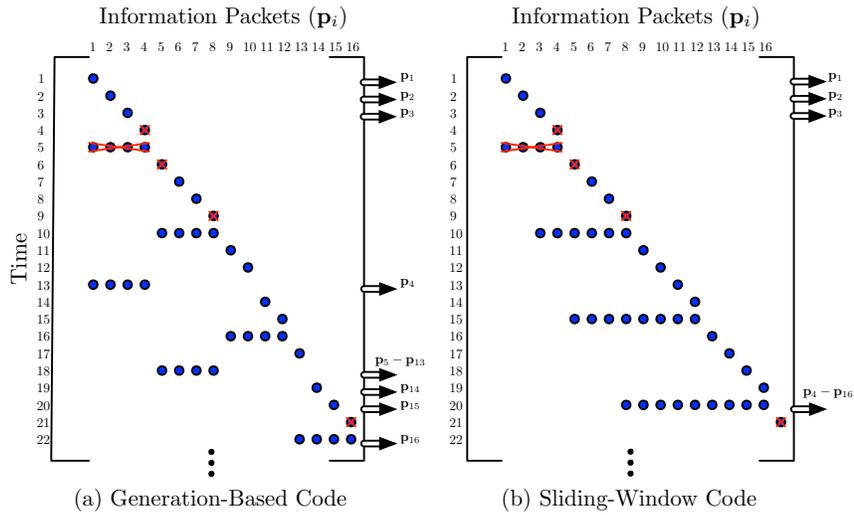}
\par\end{centering}

\caption{Examples of generation-based and sliding-window network coding
schemes. It is important to note that the generation-based coding scheme
requires feedback and retransmissions to ensure reliable delivery while the
sliding-window coding scheme only requires feedback to help slide the coding
window.}
\label{fig:NC_Examples}
\end{figure}

Examples of both schemes are provided in Figure \ref{fig:NC_Examples}. Columns
within the figure represent information packets that need to be sent, rows
represent the time when a specific packet is transmitted, and the elements of
the matrix indicate the composition of the transmitted packet. For example,
packet $\mathbf{p}_1$ is transmitted in time-slot $1$, while coded packet
$\mathbf{c}_5=\sum_{i=1}^4 \alpha_i \mathbf{p}_i$ is transmitted in time-slot
$5$. The double-arrows on the right of each matrix indicate when an information
packet is delivered, in-order, to an upper-layer application, and the red
crosses mark lost packets.

Each approach has its benefits and drawbacks. It is easy from a coding
perspective to implement the generation-based coding scheme, and these
schemes achieve capacity when $k\rightarrow \infty$. However, partitioning
packets into generations adds artificial restrictions on the code's capability
to recover from losses, and may not be as efficient as sliding window
schemes. Furthermore, generation-based schemes can increase the complexity of
the feedback process, especially for reliable data transfers.
Sliding-window schemes, on the other hand, can outperform generation-based
schemes in terms of efficiency and delay. Unfortunately, coding window
management can be difficult and these schemes typically cannot guarantee
a decoding event occurs before the termination of a session. In addition, the
size of the coding window maybe much larger than the generation-based schemes
leading to increased decoding complexity and communication overhead.

The examples shown in Figure \ref{fig:NC_Examples} will be used throughout the remainder
of this paper in order to provide some intuition into the trade-offs of using
one type of coding approach over the other. Algorithm 
\ref{alg:Generation-Coding-algorithm} describes the packet generation policy for
the generation-based scheme shown in Figure \ref{fig:NC_Examples}(a), while Algorithm 
\ref{alg:Sliding-window-coding-algorithm} describes the policy for the
sliding-window coding scheme shown in Figure \ref{fig:NC_Examples}(b). Each
algorithm uses a systematic approach where information packets $\mathbf
{p}_i,i\in\mathcal{P}$, are first sent uncoded and redundancy is added to help  
correct packet erasures by inserting coded packets into the packet stream. We  
will assume that the amount of redundancy added to the packet stream is defined 
by $R\geq 1$ (e.g., the code rate is $c=\nicefrac{1}{R}$).
\begin{figure}
\begin{minipage}[t]{0.5\textwidth}
	\vspace{0pt}
	\begin{algorithm}[H]
		\SetAlgoNoLine
		\SetAlgoNoEnd
		\DontPrintSemicolon
		\SetKwFor{ForEach}{for each}{do}{endfch}

		\ForEach{$j\in \left[ 1,\lceil \frac{\left|\mathcal{P}\right|}{k} \rceil
		\right]$}{
			$w_l \leftarrow \left(j-1\right)k+1$\;
			$w_u \leftarrow \min\left(jk,\left|\mathcal{P}\right|\right)$\;
			\ForEach{$i\in \left[ w_l,w_u\right]$}{
				Transmit $\boldsymbol{p}_i$\;
			}
			\ForEach{$m\in \left[ 1,k\left( R-1 \right) \right]$}{
				Transmit $\boldsymbol{c}_{j,m}=\sum_{i=w_l}^{w_u}
				\alpha_{i,j,m}\boldsymbol{p}_i$\;
			}
		}

		\caption{Generation-based coding algorithm \cite{Cloud_CodedGenSRARQ_2015}}
		\label{alg:Generation-Coding-algorithm}
	\end{algorithm}
\end{minipage}
\begin{minipage}[t]{0.5\textwidth}
	\vspace{0pt}
	\begin{algorithm}[H]
		\SetAlgoNoLine 
		\SetAlgoNoEnd 
		\DontPrintSemicolon
		\SetKwFor{ForEach}{for each}{do}{endfch} 

		Initialize $k=1$, $u=1$, and $n=\frac{R}{R-1}$\; 
		\ForEach{$k\in \mathcal{P}$}{    
			\If{$u < n$}{ 
				Transmit packet $\boldsymbol{p}_k$\;         
				$u\leftarrow u +1$\;            
			}     
			\Else{         
				Transmit $\boldsymbol{c}_{k}=\sum_{i=1}^{k} \alpha_{k,i}
				\boldsymbol{p}_i$\;
				$u\leftarrow 1$\;     
			}
		}
		\vspace{1pt}

		\caption{Sliding window coding algorithm \cite{Karzand_LDRLC_2014}}
		\label{alg:Sliding-window-coding-algorithm}
	\end{algorithm}
\end{minipage}
\end{figure}

It is important to note that feedback is not addressed in these algorithms. In
general, feedback is necessary to accurately estimate the network packet erasure
rate. Furthermore, feedback maybe required to ensure reliable delivery in some
instances. For the generation-based scheme, the server may need to know the
number of received degrees of freedom from each transmitted generation.
This feedback can be used by the server to retransmit additional degrees
of freedom if a particular generation cannot be decoded. Details are provided in
\cite{Cloud_CodedGenSRARQ_2015}. In the sliding window scheme, knowledge of the
number of received degrees of freedom may not be
necessary \cite{Karzand_LDRLC_2014}; but feedback can be used to help
slide the coding window or facilitate decode events if there are delay
constraints.

\section{Network Coding Performance for Packet Streams}
\label{sec:performance_of_nc}

As we mentioned in the previous section, we will compare the performance of two
types of intra-session network coding schemes (see Algorithms
\ref{alg:Generation-Coding-algorithm} and
\ref{alg:Sliding-window-coding-algorithm}) for both a reliable data stream
(e.g., a TCP session) and an unreliable data stream (e.g., a UDP session). The
metrics used to evaluate both coding schemes will depend slightly on the type of
data stream; however, the following definitions will be used throughout this section.

\begin{definition}
	The in-order delivery delay $D$ is the difference between the time an
	information packet is first transmitted and the time that the same packet is
	delivered, in-order.
	\label{def:delay}
\end{definition}

\begin{definition}
	The efficiency $\eta$ of a coding scheme is defined as the total number of
	degrees of freedom (i.e., the total number of information packets) that need
	to transfered divided by the actual number of packets (both uncoded and coded)
	received by the sink.
	\label{def:efficiency}
\end{definition}

Both of these metrics are particularly important for satellite communication
systems. In the case of reliable data streams, large propagation delays can
compound the effects of packet losses by creating considerable backlogs
and in-order delivery delays. For large file transfers or non-time sensitive
applications, this may not be an issue. However, a large number of
time-sensitive applications (e.g., non-real-time video streaming) use TCP. 
Lost packets can result in very large resequencing delays that can seriously
degrade the quality of user experience. Network coding is particularly useful in
these situations to help recover from packet losses without excessive
retransmissions. Furthermore, bandwidth is expensive for these systems. Any
coding scheme that promises to provide a specified quality of service needs to
be efficient. 

The remainder of this section will provide an outline of the network
model and examine the performance of the two coding schemes presented above. The
two metrics defined earlier will be used in addition to any additional metrics
that are important for the specific type of data stream. 

\subsection{Network Model}
\label{subsec:Network_Model}

We will assume a time-slotted model where
each time-slot has a duration $t_s$ equal to the time it takes to transmit a
single packet. The network propagation delays will be taken into account by
defining $t_p = \nicefrac{RTT}{2}$ where $RTT$ is the round-trip time. As a
reminder, we will assume that the amount of redundancy added ($R\geq
\nicefrac{1}{1-\epsilon}$ given that $\epsilon$ is the packet erasure
probability) defines the code rate $c=\nicefrac {1}{R}$. For
the generation-based scheme, $c$ is equal to the generation size divided by the
number of degrees of freedom transmitted for that generation (i.e.,
$c=\nicefrac{k}{Rk}$ where $k$ is the generation size). In the case of the
sliding window scheme, $c$ is dependent on the number of consecutively
transmitted information packets (i.e., $c=\nicefrac{n-1}{n}$ where
$n=\nicefrac{R}{R-1}$ is the number of packets between each inserted coded
packet).

The satellite channel will be modeled using a simple Gilbert channel with
transition probability matrix
\begin{equation}
	P = \begin{bmatrix}
			1-\gamma & \gamma \\
			\beta & 1-\beta
		  \end{bmatrix}
\end{equation}
where $\gamma$ is the probability of transitioning from the \enquote{good} state
(which has a packet erasure rate equal to zero) to the \enquote{bad} state
(which has a packet erasure rate equal to one) and $\beta$ is the probability of
transitioning from the \enquote{bad} state to the \enquote{good} one. The
steady-state distribution of the \enquote{bad} state $\pi_B = \nicefrac{\gamma}
{\gamma+\beta}$
and the expected number of packet erasures in a row $\mathbb{E}
\left[L\right]=\nicefrac{1}{\beta}$ will be used as the primary parameters for
determining the transition probabilities of the channel model. It should be
noted that this model does not necessarily reflect the effects of fading due to
scintillation or rain, which generally have a duration equal to hundreds of
milliseconds to hours. Instead, the model is intended to help model the cases
where the SNR is such that the performance of the underlying physical layer code
is degraded; but the situation does not warrant the need to change to a more
robust modulation/coding scheme. 
\vspace{0.5cm}
{\setlength{\parindent}{0cm}
\fbox{
	\parbox{\textwidth}{Lesson Learned: Network coding is not a cure-all solution.
It cannot mitigate \par the effects of deep fades with very large durations.}}}

\subsection{Reliable Data Stream Performance}
\label{subsec:Reliable_Streams}

Reliable data delivery is a fundamental requirement for some applications.
This section will focus on the performance of both a generation-based and a
sliding-window coding scheme by looking at the following metrics:
the ability of the scheme to provide $100\%$ reliability, the in-order delivery
delay, and the coding schemes' efficiency. Furthermore, the performance of an
idealized version of selective-repeat ARQ will be provided to highlight the
gains network coding can provide in satellite communications systems.

Before proceeding, feedback maybe necessary to ensure
reliability. With regard to the two example coding schemes presented here, the
generation-based scheme requires feedback while the sliding-window scheme does
not. Algorithm \ref{alg:Generation-Coding-algorithm} can be modified to
include this feedback with only a few changes. Assume that delayed feedback
contains information regarding the success or failure of a specific generation
being decoded by the client. If a decoding failure occurs, the server can then
produce and send additional coded packets from that generation to overcome the
failure. On the other hand, the construction of the sliding-window scheme
outlined in Algorithm \ref{alg:Sliding-window-coding-algorithm} has been shown
in \cite{Karzand_LDRLC_2014} to provide a finite in-order delivery delay with
probability one. Therefore, our results will assume that no feedback is
available when using this scheme even though feedback may actually increase the
algorithm's
performance.

A detailed analysis of the in-order delivery delay and the efficiency for the
generation-based scheme ($\mathbb{E}\left[D_G\right]$ and $\eta_G$
respectively) is provided in \cite{Cloud_CodedGenSRARQ_2015}, while the
same is provided in \cite{Karzand_LDRLC_2014} for the sliding-window scheme
($\mathbb{E}\left[D_S\right]$ and $\eta_S$ respectively).
The analysis of the generation-based scheme shows that
$\mathbb{E}\left[ D_G \right]$ and the delay's variance $\sigma^2_G$ are
dependent on both the generation size $k$ and the amount of added redundancy
$R$. For a given $R$ that is large enough and independent and identically
distributed (i.i.d.) packet losses, $\mathbb {E}
\left[ D_G\right]$ is convex with respect to $k$ and has a global minimum.  
Determining this minimum, $\mathbb{E} \left[ D_G^*\right]=\arg \min_k \mathbb{E}
\left[ D_G\right]$, is difficult due to the lack of a closed form expression;  
however it can be found numerically. The following results will only show
$\mathbb{E} \left[D_G^*\right]$ for a given $R$ since the behavior of $\mathbb{E} \left[
D_G\right]$ and $\sigma_G^2$ as a function of $k$ is provided in
\cite{Cloud_CodedGenSRARQ_2015}. The analysis of the sliding-window
scheme's in-order delivery delay shows that $\mathbb{E}\left[ D_S\right]$ is
only dependent on $R$ since there is no concept of generation or block size.
Therefore, a simple renewal process can be defined and a lower-bound for the
expected in-order delay can be derived. While the efficiency of this scheme is
not explicitly given in \cite{Karzand_LDRLC_2014}, it can easily be shown that
the efficiency is $\eta_S=\nicefrac {1} {R \left(1-\epsilon\right)}$
for i.i.d. packet losses that occur with probability $\epsilon$. Regardless of this existing analysis, the in-order
delivery delay and efficiency used below for both the generation-based and
sliding-window coding schemes are found using simulations developed in Matlab.

\begin{figure}
	\centering
	\includegraphics[width=1\textwidth]{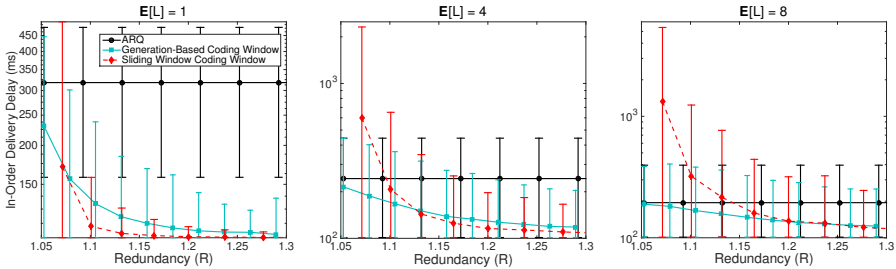}
	\caption{In-order packet delay ($\mathbb{E}\left[D\right]$) as a function of
	the redundancy ($R$) where $RTT = 200$ ms, $t_s = 1.2$ ms, and $\pi_B=0.05$.}
	\label{fig:Reliable_Delay}
\end{figure}

\begin{figure}
	\centering
	\includegraphics[width=1\textwidth]{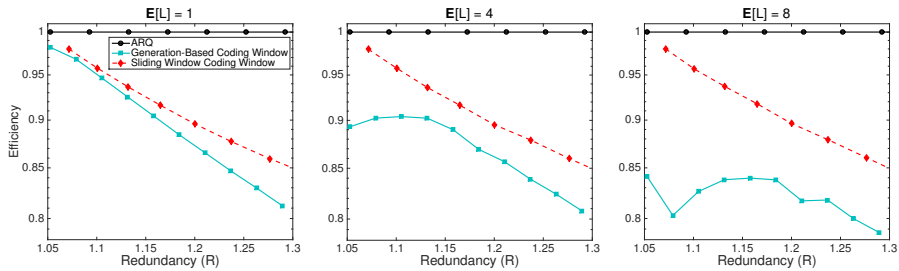}
	\caption{Efficiency ($\eta$) as a function of the redundancy ($R$) where   
	$RTT = 200$ ms, $t_s = 1.2$ ms, and $\pi_B=0.05$.}
	\label{fig:Reliable_Efficiency}
\end{figure}

Figures \ref{fig:Reliable_Delay} and \ref{fig:Reliable_Efficiency} show
$\mathbb{E}\left[D\right]$ and $\eta$ respectively for both coding schemes as a 
function of $R$. Furthermore, each sub-figure shows the impact correlated losses
have on the schemes' performance where $\mathbb{E}\left[L\right]$ is the
expected
number of packet losses that occur in a row. For uncorrelated losses (e.g.,
$\mathbb{E}\left[L\right]=1$), both coding schemes provide an in-order delivery
delay that is superior to the idealized version of selective repeat ARQ. This
performance gain becomes less pronounced as $\mathbb{E}\left[L\right]$
increases. In fact, the sliding-window coding scheme performs worse than ARQ for
small $R$ when $\mathbb{E}\left[L\right]=8$. The cause of this is due to the
lack of feedback, which can help overcome the large number of erasures if it is
implemented correctly. Regardless, Figure \ref{fig:Reliable_Delay} shows
that coding can help in the cases where losses are correlated; although the
gains come with a cost in terms of efficiency.

\vspace{0.5cm}
{\setlength{\parindent}{0cm}
\fbox{
	\parbox{\textwidth}{Lesson Learned: While feedback is necessary for
	estimating the channel/network state, it also aids in decreasing
	in-order delivery delay.}}}
\vspace{0.5cm}


Decreasing $\mathbb{E}\left[D\right]$ results in
decreased $\eta$, which can be observed in Figure \ref{fig:Reliable_Efficiency}.
The figure shows that the sliding-window coding scheme is more efficient than
the generation-based scheme. There are two major contributors to this
behavior. First, code construction has a major impact on efficiency.
Since coding occurs over more information packets in the sliding-window
scheme, coded packets can help recover from packet erasures that occur over a
larger span of time (i.e., multiple generations if we compare it with the
generation-based scheme). Second, the decrease in the generation-based scheme's
efficiency, as well as the non-decreasing behavior of $\eta_G$, for $\mathbb
{E}\left[L\right]>1$, is an indication that retransmissions are necessary to
provide reliability. In fact, the generation-based scheme almost always requires
retransmissions to be made when $\mathbb{E}\left[L\right]=8$. This behavior
helps illustrate that artificially restricting the coding window's size can have
negative impacts and may not be the appropriate strategy in certain
circumstances.

\vspace{0.5cm}
{\setlength{\parindent}{0cm}
\fbox{
	\parbox{\textwidth}{Lesson Learned: Generation-based coding schemes perform
	poorly when packet losses are correlated due to the limited number of packets
	that are used to form a coded packet.}}} 

\subsection{Unreliable Data Stream Performance} 
\label{subsec:Unreliable_Streams}

Data streams such as real-time voice and video do not necessarily require
$100\%$ reliability. However, decreasing the underlying packet erasure rates may
still drastically improve upper layer quality of service. Recent
work in this area has shown that network coding is one tool that can help
improve performance \cite{Teerapittayanon_NCoverWimax,7037035}.
This section will compare both the generation-based and sliding-window coding
schemes with respect to the upper-layer packet erasure probabilities and the
expected in-order delivery delays. 

The generation-based coding scheme shown in Algorithm
\ref{alg:Generation-Coding-algorithm}, where feedback is only necessary to
identify the packet erasure rate, is ideally suited to the case where
there is a delay constraint and packet delivery is not guaranteed. Packets within
each generation are delivered in-order until the first packet loss is
encountered. Once the entire generation has been received, the client attempts
to decode it. If the generation cannot be decoded, only the successfully
received information packets are delivered. If the generation can be decoded,
every information packet contained in the generation is delivered in-order.

Modifying the sliding-window coding scheme shown in Algorithm
\ref{alg:Sliding-window-coding-algorithm} for unreliable data streams is
somewhat difficult. If a delay constraint exists, the coding window cannot be
arbitrary changed to accommodate these constraints. For example, assume that a 
lost information packet $\mathbf{p}_i$ is no longer necessary due to its
delivery time exceeding some specified value. One approach would be to move the
left side of the coding window to the right so that $\mathbf{p}_i$ is no
longer used in the generation of future coded packets (i.e., $\mathbf
{c}_j =\sum_{k=i+1}^j \alpha_{j,k}\mathbf{p}_k$). In order for these new coded
packets to be useful, the decoder must discard any coded packet containing
$\mathbf{p}_i$ that it has already received. Not only does this decrease the  
efficiency of the coding scheme, but it also potentially increases
the delay for subsequent packets $\mathbf{p}_j,i<j$. As a result, we will
assume that Algorithm \ref{alg:Sliding-window-coding-algorithm} is left
unchanged in this scenario.

\vspace{0.5cm}
{\setlength{\parindent}{0cm}
\fbox{
	\parbox{\textwidth}{Lesson Learned: Great care must be taken when modifying
	a sliding-window coding schemes' coding window when trying to meet a delay
	constraint. Not doing so properly can lead to decreased efficiency and
	increased in-order delivery delay for subsequent packets.}}} 
\vspace {0.5cm}

\begin{figure} 
	\centering
	\includegraphics[width=1\textwidth]{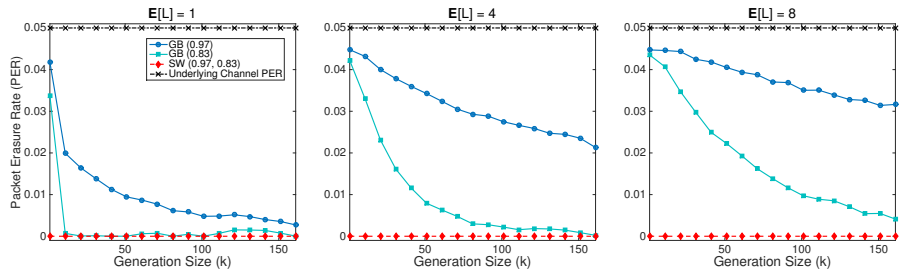}
	\caption{Upper layer packet erasure rate ($PER$) as a function of the
	generation-based coding scheme's generation size ($k$) where $RTT = 200$ ms,
	$t_s = 1.2$ ms, and $\pi_B=0.05$. The values shown within the parentheses for
	each item in
	the legend indicate the efficiency $\eta$.}
	\label{fig:Unreliable_PER}
\end{figure}

\begin{figure}
	\centering
	\includegraphics[width=1\textwidth]{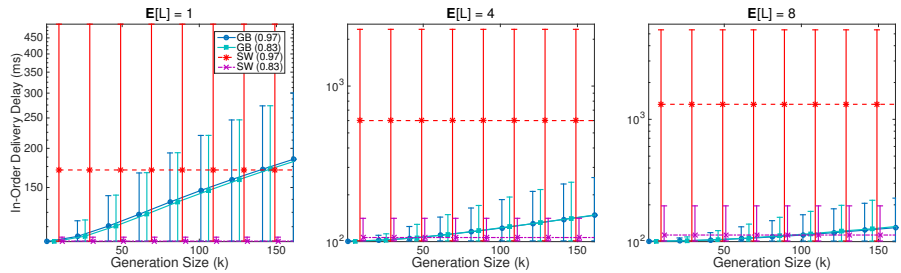}
	\caption{In-order delivery delay $\mathbb{E}\left[D\right]$ as a function of
	the generation-based coding scheme's generation size ($k$) where $RTT = 200$ ms,
	$t_s = 1.2$ ms, and $\pi_B=0.05$. The error bars
	show two standard deviations above and below the mean. The values shown within
	the parentheses for each item in
	the legend indicate the efficiency $\eta$.}
	\label{fig:Unreliable_Delay}
\end{figure}

Figures \ref{fig:Unreliable_PER} and \ref{fig:Unreliable_Delay} show the
expected upper-layer packet erasure rate ($PER$) and expected in-order
delivery delay $\mathbb{E}\left[D\right]$ respectively for both
the generation-based (GB) and sliding-window (SW) coding schemes. Three values
of the expected number of packet losses in a row $\mathbb{E}\left[L\right]$ and
two levels of efficiency $\eta$ (indicated by the values shown in parentheses)
are provided. Due to the sliding-window coding scheme's construction, the
$PER$ and $\mathbb{E}\left[D_S\right]$ are constant with respect to $k$.

These figures illustrate some of the trade-offs that need to be taken into
account when selecting the appropriate code. First, the larger the generation
size in the generation-based scheme, the better the error performance. This is
expected since you are essentially averaging losses over more packets.
However, the cost is increased latency. Second, correlated losses can have a
significant impact on the performance of the generation-based code. This is a
result of partitioning information packets into
generations, which places artificial constraints the ability of the code to
recover from packet losses. The sliding-window scheme has no such constraints. On
the other hand, the redundancy inserted into the packet stream must be enough to
ensure that any delay constraints are satisfied. For example, Figure
\ref{fig:Unreliable_Delay} shows that $\mathbb{E}\left[D_S\right]$ and
$\sigma_S$ can be very large if your goal is to be highly efficient (e.g.,
$\eta_S \approx 0.97$). In order to match the delay of the generation-based
code, a significant amount of redundancy must be added to the packet stream.

\vspace{0.5cm}
{\setlength{\parindent}{0cm}
\fbox{
	\parbox{\textwidth}{Lesson Learned: Decreasing the efficiency of
	sliding-window coding schemes is necessary to outperform generation-based
	schemes in terms of in-order delivery delay.}}} 


\section{Implementation Considerations}
\label{sec:Implementation_Considerations}

Implementing any type of network coding scheme presents its own challenges.
Sections \ref{sec:Implementing_NC} and \ref{sec:performance_of_nc} highlighted
just a few of them. However, there are a number of items that also affect how we
code, especially in satellite networks. While we cannot address everything, we  
do provide a brief discussion on some of the items that we believe are
important.

\begin{figure}
	\centering
	\includegraphics[width=0.75\textwidth]{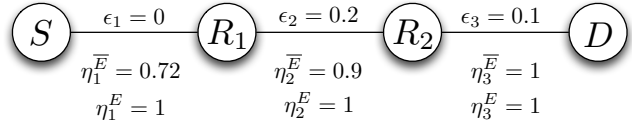}
	\caption{A simple example showing that coding within the network is more
	efficient than end-to-end coding. $\eta_i^j$ is the efficiency on link $i\in
	{1,2,3}$ when coding is performed end-to-end ($j=\bar{E}$) or at each
	intermediate network node ($j=E$).} \label{fig:RecodingExample}
\end{figure}

The first major consideration is where to perform the coding and decoding
operations. Ideally, redundancy should be added at any point in the
network where packet losses occur. This includes locations such as queues
or links where the physical layer cannot provide $100\%$ reliability.
Furthermore, the amount of added redundancy should only be enough to help
recover from losses that occur between network nodes that can code. This can be
motivated by the simple example shown in Figure \ref{fig:RecodingExample} where
a source $S$ wants to transmit $\left|\mathcal{P}\right|$ packets to the  
destination $D$. However, these packets must travel over a tandem network where
each link $i\in\{1,2,3\}$ has an i.i.d. packet erasure probability $\epsilon_i$.
If end-to-end coding is used, $\left|\mathcal{P}\right|\left(\prod_i
\left(1-\epsilon_i\right)^{-1} - 1\right)$ coded packets must be generated at
$S$ and transmitted through the network. This results in an inefficient use of
links closer to the source than would be necessary if redundancy is included
into the packet stream at each node $R_i,i\in{1,2}$.

\vspace{0.5cm}
{\setlength{\parindent}{0cm}
\fbox{
	\parbox{\textwidth}{Lesson Learned: Coding at intermediate nodes, rather than
	coding end-to-end increases	overall network efficiency.}}}
\vspace{0.5cm}

This simple fact can have major implications for satellite networks since
bandwidth is limited and very expensive. As a result, coding should be
performed at each satellite gateway or performance enhancing proxy (PEP) at a
minimum; and if possible, at each hop in the satellite network. While coding
should be performed as often as possible, network codes do not need to be
decoded at each hop. This is also extremely beneficial in satellite networks
since you can essentially shift a large portion of the required processing to
the satellite gateway or end client. In other words, coded packets can be
generated at multiple points within the network while only needing to decode
once at the client or satellite network gateway. In the example provided in
Figure \ref{fig:RecodingExample}, coding can take place at $S$, $R_1$, and
$R_2$; however, only $D$ needs to decode.

\vspace{0.5cm}
{\setlength{\parindent}{0cm}
\fbox{
	\parbox{\textwidth}{Lesson Learned: Decoding only needs to be performed once
	regardless of the number of times coding occurs within the network.}}}
\vspace{0.5cm}

The second consideration that needs to be taken into account is how to
communicate the coding coefficients $\alpha_i$ used to the decoder. For
generation-based coding schemes where $k$ is typically small, one can simply
insert each coding coefficient into the header, which would require $qk$ bits
assuming each $\alpha_i \in \mathbb{F}_{2^q}$. Coding
within the network only needs to modify the existing coefficients and does not
increase the size of the coding coefficient vector. Of course, other approaches
that require less than $qk$ bits such as \cite{Lucani_Fulcrum} or \cite{6324374}
can be used to decrease overhead. 

Communicating the coefficients efficiently for sliding-window schemes is more
challenging since the coding windows can be quite large.
Existing methods typically use a pseudo-random number generator and
communicate only the seed. This seed is then used by the decoder to generate  
the coefficients used to create each coded packet. Unfortunately, this does not 
scale well when coding occurs at intermediate network nodes. As an
example, assume that an intermediate node's coding window contains multiple
coded packets that were generated by previous nodes. When the node generates a  
new coded packet, it must communicate the seed used to generate the packet; in  
addition to all of the seeds for each of the coded packets contained within its
coding window. If the coding window and the number of coded packets contained
within the window are large, the amount of overhead required to reproduce the
coefficients can far exceed the payload size.

\vspace{0.5cm}
{\setlength{\parindent}{0cm}
\fbox{
	\parbox{\textwidth}{Lesson Learned: The overhead required to communicate
	coding coefficents for sliding-window based schemes can be significant if
	not done correctly.}}} 
\vspace{0.5cm}

Finally, congestion control and file size can potentially dictate the coding
approach used. Regardless of the type of data stream, some form of congestion
control is typically needed at either the client/server or at the satellite
network gateway. Common congestion control algorithms can cause bursts of
packets, or packet trains, while they are ramping up to fully utilize the
network. This behavior is even more pronounced when considering TCP flows over
satellite networks. In these situations, it maybe preferable to use a coding
scheme that provides a high probability of delivering every packet within a
burst without needing retransmissions or waiting for the next packet burst to
arrive. For example, a generation-based coding scheme can be used for small
congestion window sizes and a sliding-window scheme can be used for large
ones.

In a similar fashion, the coding strategy can also significantly impact the
overall throughput for some file sizes. For example, consider a small file that
can be transmitted using less than a single bandwidth-delay product worth of
packets. A generation-based coding scheme, or a mixture of the generation-based
and sliding-window schemes, should be used so that the the probability of
decoding the file after the first transmission attempt is made very large. While
this may impact the efficiency of the network, it can have major benefits
for the user's quality of service or experience.

\vspace{0.5cm}
{\setlength{\parindent}{0cm}
\fbox{
	\parbox{\textwidth}{Lesson Learned: Congestion control and the length of the
	data stream may affect the network coding strategy.}}}
	
\section{Conclusion}
\label{sec:Conclusion}

Intra-session network coding is a promising technique that can help improve
application layer performance in challenging space-based data packet networks.
However, implementing it can be problematic if done incorrectly.
This paper used two common examples of intra-session network codes to show
the benefits and drawbacks of one over the other. The first example used was a
generation-based network code and the second a sliding-window based network
code. While generation-based network codes are easier to implement,
sliding-window network codes can provide improved performance in terms of
in-order delivery delay and efficiency. This is especially the case when
reliability is required. However, generation-based network codes are able to
provide strict delay guarantees and improved upper layer packet erasure rates
with little impact to the overall network efficiency when reliability is not a
constraint. On the other hand, implementation considerations typically limit the
performance of sliding-window network codes in these environments.

Lessons learned, as well as other implementation tips, were provided in
addition to the above comparison. Some of the more important lessons learned
include the facts that restricting the size of the coding window in any way
limits the network code's performance gains; and feedback is useful for not
only estimating the channel/network state information, but it also can be used
to decrease delay. Both of these are apparent when considering the effects
correlated packet losses have on the delay for reliable data streams. Various
implementation considerations were also highlighted. These include where
coding and decoding within the network should occur, how congestion control
affects the way we code, and the challenges regarding the communication of
RLNC coefficients between the source and sink. While properly implementing
network coding in real networks can be difficult, we hope that our lessons
learned will aid in the deployment of network codes in future satellite
communication systems.

\bibliographystyle{IEEEtran}
\bibliography{NCforSATCOM}

\end{document}